\documentclass[a4paper,11pt]{article}

\usepackage{amssymb,amsmath}

\begin{document}

\title{\bf{A Paradox on Quantum Field Theory of Neutrino Mixing and Oscillations}}

\author{
{\large Y.F. Li and Q.Y. Liu}
\bigskip
\\
  {\small \it Department of Modern Physics, University of Science
    and}
\\
{ {\small \it  Technology  of China, Hefei, Anhui 230026, China.}
}
}

\date{}
\maketitle \vskip 12mm

\begin{abstract}
Neutrino mixing and oscillations in quantum field theory framework
had been studied before, which shew that the Fock space of flavor
states is unitarily inequivalent to that of mass states
(\emph{inequivalent vacua model}). A paradox emerges when we use
these neutrino weak states to calculate the amplitude of $W$ boson
decay. The branching ratio of $W^+ \rightarrow e^++\nu_\mu$ to
$W^+ \rightarrow e^++\nu_e$ is approximately at the order of
$O({m_i^2}/{k^2})$. The existence of flavor changing currents
contradicts to the Hamiltonian we started from, and the usual
knowledge about weak processes. Also, negative energy neutrinos
(or violating the principle of energy conservation) appear in this
framework. We discuss possible reasons for the appearance of this
paradox.

\end{abstract}

\vspace{3mm}

\vskip 12mm

{\small PACS numbers: 14.60.Pq, 14.60.Lm, 13.38.Be

Keywords: Neutrino mixing; flavor neutrino; W boson decay}

\vskip 5cm {\large \textbf{ 
}} \vfill \eject
\baselineskip=0.30in
\renewcommand{\theequation}{\arabic{section}.%
\arabic{equation}} \renewcommand{\thesection}{\Roman{section}}
\makeatletter
\@addtoreset{equation}{section} \makeatother

\section{Introduction}
Neutrino oscillation
experiments\cite{sno,Kamiokande,SK,kldet,CHOOZ,K2K} give
compelling evidences for neutrino oscillation theory. But there
are some difficulties in theoretical aspects about the mixing
fields in Quantum Field Theory (QFT), such as the definition of
weak states\cite{kim92,kimbook}, or equivalently the definitions
of the
operators for creating and annihilating a weak state particle.\\
The \emph{inequivalent vacua
model}\cite{BV95,BV98,BV02,remarks,space,BV05} is constructed with
a preceding attitude. In this model the transformation between
Fock space of mass states and flavor states is a \emph{bogliubov}
transformation. Basic results of this model are: unitary
in-equivalence between mass vacuum and flavor vacuum; fermion
condensation in vacuum responsible for correction to the usual
oscillation formulas and so on. An exact neutrino oscillation
formula is obtained there, which leads the usual Pontecovo's
oscillation formula to an approximate convenience.\\
In this model, there is freedom to choose spinors to expand the
flavor fields $\nu_{\sigma}(x)$. We can use a series of spinors
\{$u_{\sigma}({\bf k},r)$,$v_{\sigma}({\bf k},r)$\}\cite{FHY,FHY2},
which satisfy free Dirac equations
\begin{equation}
(\not\!k-\mu_{\sigma})u_{\sigma}({\bf k},r)=0\,,
\end{equation}
\begin{equation}
(\not\!k+\mu_{\sigma})v_{\sigma}({\bf k},r)=0\,,
\end{equation}
where $\mu_{\sigma}$ are free mass parameters. This degree of
freedom implies that we have infinite equivalent Fock space. In
respect that, the author of ref.\cite{Giunti03} thinks that the
arbitrary parameters $\mu_{\sigma}$ can be physical observables,
so he argues that Fock states of flavor neutrinos are unphysical
\cite{Giunti03,Giunti04}. But the authors in
ref.\cite{remarks,BV05} demonstrate that the oscillation formulas
in vacuum are free from the arbitrariness of the mass parameter
$\mu_{\sigma}$. So we will omit this problem in this paper, and
use their initial expansions. Our focus is to study weak processes
in this \emph{inequivalent vacua model}. The results come out that
a paradox appears even if we
carry out everything correctly.\\
The paper is organized as follows: in section II, we give the
basic aspects of the \emph{inequivalent vacua model}\,; in section
III, we will derive our calculations for $W$ boson decay and give
our main results of this paper; in section IV, we give the
conclusions and comments.

\section{Basic aspects of the \emph{inequivalent vacua model}}
Following the previous study of the neutrino mixing in QFT
\cite{BV95,BV98,BV02,remarks,space,BV05,FHY,FHY2}, In this section
we start our derivations in a two-generation case, and will give
general formulas for $N$ generations at the end of this section
which are useful in our main calculations in this paper. The
\emph{bogliubov} transformation is defined as
\begin{eqnarray}
\left(\begin{array}{c}
\nu_e(x)\\
\nu_\mu(x)\\
\end{array}\right)&=&G^{-1}(\theta;t)
\left(\begin{array}{c}
\nu_1(x)\\
\nu_2(x)\\
\end{array}\right)G(\theta;t)
\nonumber\\
&=&\left(\begin{array}{rr}
\cos\theta & \sin\theta\\
-\sin\theta & \cos\theta\\
\end{array}\right)\left(\begin{array}{c}
\nu_1(x)\\
\nu_2(x)\\
\end{array}\right)\,.
\end{eqnarray}
$G(\theta;t)$ is given by
\begin{equation}
G(\theta;t){=}exp\{\theta\int
d^{3}{\bf x}[\nu^+_1(x)\nu_2(x)-\nu^+_2(x)\nu_1(x)]\}\,,
\end{equation}
where $t=x_0$\,, \{$\nu_\sigma(x),\sigma=e,\mu$\} and
$\{\nu_i(x),i=1,2\}$ are the neutrino fields with definite flavors
and masses, respectively.

\noindent The mass fields are expanded as
\begin{eqnarray}
\nu_i(x)&=&\frac{1}{(2\pi)^{3/2}}\sum_{r}\int d^{3}{\bf k}
[u_i({\bf k},r)a^r_{{\bf k},i}e^{-i\omega_it}{+}
v_i(-{\bf k},r)b^{r^\dag}_{-{\bf k},i}e^{i\omega_it}]e^{i {\bf k} \cdot {\bf x}}
\nonumber\\
&\equiv&\frac{1}{(2\pi)^{3/2}}\sum_{r}\int d^{3}{\bf k}
[u_i({\bf k},r)a^r_{{\bf k},i}(t){+}
v_i(-{\bf k},r)b^{r^\dag}_{-{\bf k},i}(t)]e^{i {\bf k} \cdot {\bf x}}\,,
\end{eqnarray}
where $\omega_i{=}\sqrt{{\bf k}^2{+}m_i^2}$\,, $u_i({\bf k},r)$
and $v_i(-{\bf k},r)$ are the solutions of free Dirac equations in
momentum space with definite spin ${r}$ and mass ${m_{i}}$:
\begin{equation}
(\not\!k-m_{i})u_i({\bf k},r)=0\,,
\end{equation}
\begin{equation}
(\not\!k+m_{i})v_i({\bf k},r)=0\,.
\end{equation}
The Hilbert space of definite mass states ${\cal H}_{1,2}$ is constructed by
operators $a^r_{{\bf k},i}(t)$ and $b^{r}_{{\bf k},i}(t)$. So the mass vacuum $|0\rangle_m$ is
defined as:
\begin{equation}
\left(\begin{array}{c}
a^r_{{\bf k},i}(t)\\
b^{r}_{{-\bf k},i}(t)\\
\end{array}\right)|0\rangle_m=0\,,
\end{equation}
with normalization\, $_m\langle 0|0 \rangle_m=1$\,.\\
As discussed above, we will use the initial expansions of flavor fields in
ref.\cite{BV95,BV98,BV02}\,. The explicit forms are
\begin{equation}\label{expf}
\nu_{\sigma}(x){=}\frac{1}{(2\pi)^{3/2}}\sum_{r}\int d^{3}{\bf k}
[u_i({\bf k},r)a^r_{{\bf k},\sigma}(t){+}v_i(-{\bf k},r)b^{r^\dag}_
{-{\bf k},\sigma}(t)]e^{i {\bf k} \cdot \bf x}\,,
\end{equation}
where $(\sigma,i)$ stands for either $(e,1)$  or $(\mu,2)$.\\
Immediately we obtain
\begin{equation}
\left(\begin{array}{c}
a^r_{{\bf k},\sigma}(t)\\
b^{r^\dag}_{{-\bf k},\sigma}(t)\\
\end{array}\right)=G^{-1}(\theta;t)\left(\begin{array}{c}
a^r_{{\bf k},i}(t)\\
b^{r^\dag}_{{-\bf k},i}(t)\\
\end{array}\right)G(\theta;t)\,.
\end{equation}
The vacuum for flavor states is
\begin{equation}
|0(t)\rangle_f=G^{-1}(\theta;t)|0\rangle_m\,.
\end{equation}
Note that the vacuum$|0(t)\rangle_f$\,is time-dependent,
so do the creation and annihilation operators of flavor states.\\
The explicit matrix form for flavor operators is
\begin{equation}\label{mixtr}
\left(\begin{array}{c}
a^r_{{\bf k},e}(t)\\
a^r_{{\bf k},\mu}(t)\\
b^{r^\dag}_{{-\bf k},e}(t)\\
b^{r^\dag}_{{-\bf k},\mu}(t)\\
\end{array}\right)=
\left(\begin{array}{cccc}
c_{\theta}\rho^{\bf k}_{1,1} & s_{\theta}\rho^{\bf k}_{1,2} &
ic_{\theta}\lambda^{\bf k}_{1,1} & is_{\theta}\lambda^{\bf k}_{1,2} \\
-s_{\theta}\rho^{\bf k}_{2,1} & c_{\theta}\rho^{\bf k}_{2,2} &
-is_{\theta}\lambda^{\bf k}_{2,1} & ic_{\theta}\lambda^{\bf k}_{2,2} \\
ic_{\theta}\lambda^{\bf k}_{1,1} & is_{\theta}\lambda^{\bf k}_{1,2} &
c_{\theta}\rho^{\bf k}_{1,1} & s_{\theta}\rho^{\bf k}_{1,2} \\
-is_{\theta}\lambda^{\bf k}_{2,1} & ic_{\theta}\lambda^{\bf k}_{2,2} &
-s_{\theta}\rho^{\bf k}_{2,1} & c_{\theta}\rho^{\bf k}_{2,2} \\
\end{array}\right)
\left(\begin{array}{c}
a^r_{{\bf k},1}(t)\\
a^r_{{\bf k},2}(t)\\
b^{r^\dag}_{{-\bf k},1}(t)\\
b^{r^\dag}_{{-\bf k},2}(t)\\
\end{array}\right)\,,
\end{equation}
where $c_{\theta} \equiv \cos{\theta}$\,,\,$s_{\theta} \equiv \sin{\theta}$ and
\begin{equation}
\rho^{\bf k}_{i,j}\delta_{rs} \equiv \cos{\frac{\chi_i-\chi_j}{2}}\delta_{rs}=
u_i^\dag({\bf k},r)u_j({\bf k},s)=v_i^\dag(-{\bf k},r)v_j(-{\bf k},s)\,,
\end{equation}
\begin{equation}
i\lambda^{\bf k}_{i,j}\delta_{rs} \equiv i\sin{\frac{\chi_i-\chi_j}{2}}\delta_{rs}=
u_i^\dag({\bf k},r)v_j(-{\bf k},s)=v_i^\dag(-{\bf k},r)u_j({\bf k},s)\,,
\end{equation}
with $i,j=1,2$ and $\cot\chi_i=|{\bf k}|/m_i$\,.\\
For $N$ generations, general formulas are similar to
(\ref{mixtr}):
\begin{equation}\label{Ntr}
a^r_{{\bf k},\sigma}(t)=\sum_{j=1}^N\{U_{\sigma,j}\rho^{\bf
k}_{i,j}a^r_{{\bf k},j}(t)+ U_{\sigma,j}i\lambda^{\bf
k}_{i,j}b^{r^\dag}_{{-\bf k},j}(t)\}\,,
\end{equation}
\begin{equation}\label{Ntr2}
b^{r^\dag}_{{-\bf
k},\sigma}(t)=\sum_{j=1}^N\{U_{\sigma,j}i\lambda^{\bf
k}_{i,j}a^r_{{\bf k},j}(t)+ U_{\sigma,j}\rho^{\bf
k}_{i,j}b^{r^\dag}_{{-\bf k},j}(t)\}\,,
\end{equation}
where the pair of $(\sigma,i)$ denotes
$((e,1),(\mu,2),(\tau,3),\cdots)$, and $U_{\sigma,j}$ is the
neutrino mixing matrix if we choose the charge leptons
to be the mass eigenstates.\\
The most important aspect of the flavor operators is the fact that
anticommutations at different time are not the standard canonical
relations but more complex. We compute the related ones below(we
fix one operator at $t=0$\,, and the other at time $t$):
\begin{equation}\label{commu1}
\{a^r_{{\bf k},\sigma}(0),a^{r^\dag}_{{\bf k},\delta}(t)\}=
\sum_{l}U_{\sigma,l}U^\ast_{\delta,l}\{\rho^{{\bf k}}_{i,l}\rho^{{\bf k}}_{j,l}e^{i\omega_lt}+
\lambda^{{\bf k}}_{i,l}\lambda^{{\bf k}}_{j,l}e^{-i\omega_lt}\}\,,
\end{equation}
\begin{equation}\label{commu2}
\{a^r_{{\bf k},\sigma}(0),b^r_{{-\bf k},\delta}(t)\}=
\sum_{l}U_{\sigma,l}U^\ast_{\delta,l}\{-i\rho^{\bf {k}}_{i,l}\lambda^{{\bf k}}_{j,l}e^{i\omega_lt}+
i\lambda^{{\bf k}}_{i,l}\rho^{{\bf k}}_{j,l}e^{-i\omega_lt}\}\,,
\end{equation}
where the pairs of$(\sigma,i)$ and $(\delta,j)$ denote
$((e,1),(\mu,2),(\tau,3),\cdots)$\,. \\
When we choose another $t=0$ in (\ref{commu1}) and (\ref{commu2}),
we can get
\begin{equation}\label{commu1t}
\{a^r_{{\bf k},\sigma}(0),a^{r^\dag}_{{\bf k},\delta}(0)\}=
\sum_{l}U_{\sigma,l}U^\ast_{\delta,l}\{\rho^{{\bf
k}}_{i,l}\rho^{{\bf k}}_{j,l}+ \lambda^{{\bf k}}_{i,l}\lambda^{{\bf
k}}_{j,l}\}\,,
\end{equation}
\begin{equation}\label{commu2t}
\{a^r_{{\bf k},\sigma}(0),b^r_{{-\bf k},\delta}(0)\}=
\sum_{l}U_{\sigma,l}U^\ast_{\delta,l}\{-i\rho^{\bf
{k}}_{i,l}\lambda^{{\bf k}}_{j,l}+ i\lambda^{{\bf
k}}_{i,l}\rho^{{\bf k}}_{j,l}\}\,.
\end{equation}
We can see that for the same flavor we get the standard canonical
anticommutations such as $\{a^r_{{\bf
k},\sigma}(0),a^{r^\dag}_{{\bf k},\sigma}(0)\}=1$ and $\{a^r_{{\bf
k},\sigma}(0),b^r_{{-\bf k},\sigma}(0)\}=0\,,$ but for different
flavors the anticommutations are nonzero due the the dependence of
$\rho^{\bf k}$ and $\lambda^{\bf k}$ on the $m_i$. This flavor
changing effect gives important results of this paper.
\\
One of the consequences of this model is an exact neutrino
oscillation formula obtained, e.g., for two-neutrino case the
survival probability \cite{BV98} is
\begin{equation}
\label{exactP} P(\nu_e \rightarrow \nu_e)= 1-sin^2 2\theta
\{|U_k|^2 sin^2[\Phi^+(t)]+|V_k|^2 sin^2[\Phi^-(t)]\},
\end{equation}
here $\Phi^+(t)$ and $\Phi^-(t)$ are oscillation phases induced by
positive and negative frequency parts; $|V_k|=\sqrt{1-|U_k|^2}$
with
\begin{equation}\label{UinP}
|U_{\bf k}|^2 = {1 \over 2} \sum_{r,s} |u_2^+({\bf k},r)u_1({\bf
k},s)|^2=1-O({m_i^2 \over k^2}).
\end{equation}
When $|U_{\bf k}|^2=1$, this exact probability becomes the usual
Pontecovo formula. Corrections from \emph{inequivalent vacua
model} are at the order of $O({m_i^2 \over k^2})$.

\section{Problems of neutrino weak states}
Now we want to use the weak states defined above to derive the
amplitudes of weak interaction processes described by charge current
($CC$) and neutral current ($NC$) in \emph{Standard Model} ($SM$) of
elementary particle physics, we get some ridiculous results after
our calculations, such as negative energy neutrinos and flavor
changing currents.

\subsection{Negative energy neutrinos}
Considering neutrinos produced through $CC$ process, such as
\begin{equation}\label{ccinteract}
W^+\rightarrow e^++\nu_e\,,
\end{equation}
the Hamiltonian responsible for this production vertex is
\begin{equation}\label{cclagran}
{\cal H}=-\frac{g}{\sqrt{2}} W^+_\mu(x)J^{\mu+}_W \equiv
-\frac{g}{2\sqrt{2}} W^+_\mu(x){\bar \nu_e}(x) \gamma^\mu
(1-\gamma^5)e(x)\,.
\end{equation}
Assuming this process to take place at $t=0$, the flavor vacuum at
$t=0$ is defined as $|0\rangle_f \equiv |0(t=0)\rangle_f$; then
one $e$-neutrino state is $|\nu_e({\bf k},r)\rangle \equiv
a^{r\dag}_{{\bf k},e}(0)|0\rangle_f$; and the Hermitian
conjugation of this state is $\langle\nu_e({\bf k},r)|
\equiv\,_f\langle 0| a^r_{{\bf k},e}(0)$\,. So the amplitude at
tree level is expressed as
\begin{equation}
i{\cal M}=\langle{\nu_e}({\bf k},r) {e^+}({\bf k}_e,r_e)|\{-i\int
{d^{4}x}{\cal H}(x)\} |{W^+}({\bf k}_W,\epsilon_{\mu}) \rangle\,.
\end{equation}
Because $e(x)$ and $W^+_\mu(x)$ are both fields with definite mass
quanta, their matrix elements can be derived easily as usual
\begin{equation}
\langle 0|W^+_\mu(x)|{W^+}({\bf k}_W,\lambda)\rangle \propto
\epsilon_{\mu}({\bf k}_W,\lambda)\,e^{-i\omega_Wt+i{\bf k}_W \cdot
{\bf x}}\,,
\end{equation}
\begin{equation}
\langle {e^+}({\bf k}_e,r_e)|e(x)|0 \rangle \propto v_e({\bf
k}_e,r_e)\,e^{i\omega_et-i{\bf k}_e \cdot {\bf x}}\,.
\end{equation}
We omit trivial constants in above expressions for simplicity,
which have no influence on our results. $\epsilon_{\mu}({\bf
k}_W,\lambda)$ is the polarization vector of the $W^+$ boson,
and $v_e({\bf k}_e,r_e)$ is the spinor of positron $e^+$.\\
Subtle differences come from neutrino sector. According to the
\emph{inequivalent vacua model}\,, we must use the flavor states
to compute the matrix elements. Based on the expansion of
(\ref{expf}), we can derive that
\begin{eqnarray}
i{\cal M}&\propto& ig \delta^{(3)}({\bf k}_W-{\bf k}_e-{\bf
k})\int{d t}
\nonumber\\
&& \{\,_f\langle 0|a^r_{{\bf k},e}(0)a^{r\dag}_{{\bf
k},e}(t)|0\rangle_f\, {\bar u_1({\bf k},r)}\gamma^\mu
(1-\gamma^5)v_e({\bf k}_e,r_e)+
\nonumber\\
&& _f\langle 0|a^r_{{\bf k},e}(0)b^r_{-{\bf k},e}(t)|0\rangle_f\,
{\bar v_1(-{\bf k},r)}\gamma^\mu (1-\gamma^5)v_e({\bf
k}_e,r_e)\,\}
\nonumber\\
&&\epsilon_{\mu}({\bf
k}_W,\lambda)e^{i\omega_et}e^{-i\omega_Wt}\,. \label{amplitude1}
\end{eqnarray}
The flavor vacuum $|0\rangle_f$ is defined at $t=0$, so matrix
elements in (\ref{amplitude1}) can be expressed as
\begin{equation}
_f\langle 0|a^r_{{\bf k},e}(0)a^{r\dag}_{{\bf k},e}(t)|0\rangle_f
=\{a^r_{{\bf k},e}(0),a^{r^\dag}_{{\bf k},e}(t)\}\,,
\end{equation}
\begin{equation}
_f\langle 0|a^r_{{\bf k},e}(0)b^r_{-{\bf k},e}(t)|0\rangle_f
=\{a^r_{{\bf k},e}(0),b^r_{{-\bf k},e}(t)\}\,.
\end{equation}
Now by using the expressions (\ref{commu1}) and (\ref{commu2}), we
can get the final result of this amplitude
\begin{eqnarray}\label{ccamplitude}
i{\cal M} &\propto& ig \delta^{(3)}({\bf k}_W-{\bf k}_e-{\bf
k})\sum_i |U_{e,i}|^{2}
\nonumber\\
&& \{\,\{\,[\,{\rho^{\bf k}_{1,i}}^{2}\delta(\omega_W - \omega_e -
\omega_i)+ {\lambda^{\bf k}_{1,i}}^{2}\delta(\omega_W - \omega_e +
\omega_i)\,]
\nonumber\\
&& {\bar u_1({\bf k},r)}\gamma^\mu (1-\gamma^5)v_e({\bf k}_e,r_e)
\epsilon_{\mu}({\bf k}_W,\lambda)\,\}+
\nonumber\\
&& \{\,[\,-i\rho^{\bf k}_{1,i} \lambda^{\bf k}_{1,i}
\delta(\omega_W - \omega_e - \omega_i)+ i\lambda^{\bf
k}_{1,i}\rho^{\bf k}_{1,i} \delta(\omega_W - \omega_e +
\omega_i)\,]
\nonumber\\
&& {\bar v_1({-\bf k},r)}\gamma^\mu (1-\gamma^5)v_e({\bf k}_e,r_e)
\epsilon_{\mu}({\bf k}_W,\lambda)\,\}\,\}\,.
\end{eqnarray}
Among four parts of this amplitude, each has one $\delta$ function
about the energy, but two of them are $\delta(\omega_W - \omega_e
+ \omega_i)$\,. If it is interpreted as the conservation of
energy, then there is negative energy neutrino with
$E=-\omega_i$\,. Or contrarily, if we think neutrinos always have
positive energy, this
process will violate the principle of energy conservation. \\
In the limit of massless neutrinos, three of the four terms in
(\ref{ccamplitude}) are vanishing and leaving only the first,
which is entirely the same as the standard expression in $SM$. But
here terms with $\delta(\omega_W - \omega_e + \omega_i)$\, are
non-vanishing due to the dependence of $\rho^{\bf k}$ ,
$\lambda^{\bf k}$ and $\delta$ functions on the index $i$.\\
Entirely degenerated mass spectrum with $m_i=m\,$ can also resolve
this problem. It indicates that $\rho^{\bf k}=1\,,\lambda^{\bf
k}=0\,$ and $\omega_i=\omega\,$, so the amplitude can be
simplified as
\begin{equation}\label{ccamplitude2}
i{\cal M} \propto ig \delta^{(4)}(k_W-k_e- k) {\bar u_m({\bf
k},r)}\gamma^\mu (1-\gamma^5)v_e({\bf k}_e,r_e)
\epsilon_{\mu}({\bf k}_W,\lambda)\,,
\end{equation}
where $u_m({\bf k},r)$ is the solution of $(\not\!k-m)u_m({\bf
k},r)=0\,$,\, ${\bf k}$ and $k^0\equiv\omega=\sqrt{{\bf
k}^2{+}m^2}$ are the momentum vector and the energy of $\nu_e$
respectively. In fact, in this case there is no mixing at all,
neutrino weak eigenstates are also mass eigenstates. It is a
generalization of the case of massless neutrinos. It is mass
differences not masses that are the crucial points of this
problem. However neutrino oscillation experiments, e.g., solar and
atmospheric neutrino oscillations have confirmed the mass
differences between different neutrinos
\cite{sno,Kamiokande,SK,kldet,MSW,LMA}, thus this problem cannot
be neglected.

\subsection{Appearance of flavor changing currents}
In fact, inspired by (\ref{commu1}) and (\ref{commu2}), we know
that anticommutations for different flavors can also give nonzero
results, so there exist non-trivial flavor changing $CC$ and $NC$
matrix elements at tree level. For example, we consider process
\begin{equation}\label{odinteract}
W^+\rightarrow e^++\nu_\mu\,.
\end{equation}
When we use the Hamiltonian responsible for the standard $CC$
interactions in (\ref{cclagran}), we get the tree-level amplitude
\begin{equation}
i{\cal M}=\langle{\nu_\mu}({\bf k},r) {e^+}({\bf
k}_e,r_e)|\{-i\int {d^{4}x}{\cal H}(x)\} |{W^+}({\bf
k}_W,\epsilon_{\mu}) \rangle\,.
\end{equation}
Non-vanishing amplitudes come from the neutrino sector again.
Because anticommutations at different time such as (\ref{commu2})
are not the standard canonical relations, we get this unexpected
amplitude. The final form of the amplitude can be expressed as
\begin{eqnarray}\label{odamplitude}
i{\cal M} &\propto& ig \delta^{(3)}({\bf k}_W-{\bf k}_e-{\bf
k})\sum_i U_{\mu,i}U^\ast_{e,i}
\nonumber\\
&& \{\,\{\,[\,{\rho^{\bf k}_{2,i}}\rho^{\bf
k}_{1,i}\delta(\omega_W - \omega_e - \omega_i)+ {\lambda^{\bf
k}_{2,i}}{\lambda^{\bf k}_{1,i}}\delta(\omega_W - \omega_e +
\omega_i)\,]
\nonumber\\
&& {\bar u_1({\bf k},r)}\gamma^\mu (1-\gamma^5)v_e({\bf k}_e,r_e)
\epsilon_{\mu}({\bf k}_W,\lambda)\,\}+
\nonumber\\
&& \{\,[\,-i\rho^{\bf k}_{2,i}\lambda^{\bf k}_{1,i}
\delta(\omega_W - \omega_e - \omega_i)+ i\lambda^{\bf
k}_{2,i}\rho^{\bf k}_{1,i} \delta(\omega_W - \omega_e +
\omega_i)\,]
\nonumber\\
&& {\bar v_1({-\bf k},r)}\gamma^\mu (1-\gamma^5)v_e({\bf k}_e,r_e)
\epsilon_{\mu}({\bf k}_W,\lambda)\,\}\,\}\,.
\end{eqnarray}
In the case of entirely degenerated mass spectrum with $\rho^{\bf
k}=1\,$ and $\lambda^{\bf k}=0\,$, the total amplitude is
vanishing due to the unitary of the mixing matrix. But in general
case, besides negative energy neutrino problem, we encounter
another severe problem: the dependence of $\rho^{\bf k}$ ,
$\lambda^{\bf k}$
and $\delta$ functions on the index $i$ makes this amplitude nonzero.\\
Now let us estimate the branching ratio of this off-diagonal modes
(\ref{odinteract}) to the normal diagonal modes
(\ref{ccinteract}). In (\ref{odamplitude}) all particles are
considered as in plane waves, and there are $\delta$ functions of
energy inside the sum. For different mass eigenstates the $\delta$
functions are different, thus they can't be taken out of the sum.
Under this consideration the branching ratio will be completely
different from that in $SM$ with zero neutrino mass. However this
phenomenon is a general effect for mixing neutrino. It is a
physical limit which describes an averaged neutrino oscillation
effect, which is put as an appendix at the end of this paper. For
an usual weak process, it is finished in a limited space-time. The
energy uncertainty makes the $\delta$ function to be replaced by a
wave package profile of energy distribution (e.g., a sharp
gaussian). Different profiles with respect to $i$ entirely overlap
thus we can factorize the $\delta$ functions out of the sum.
Because in the rest frame of the $W^+$ boson, the momentum of
neutrinos almost equals to $m_W/2$ ($m_W$ is the mass of $W$
boson, approximately equals 80 GeV), which is much larger than the
masses of neutrinos. We expand the non trivial $\rho^{\bf k}$ and
$\lambda^{\bf k}$ to high orders: $\rho^{\bf k}\sim
1-O(\frac{m_i^2}{k^2})$\,, $\lambda^{\bf k}\sim
O(\frac{m_i}{k})$\,, and only consider the leading term in the two
amplitudes. The estimated branching ratio will be
\begin{equation}\label{ratio2}
R_{\nu_\mu/\nu_e}\equiv\frac{\Gamma(W^+\rightarrow
e^++\nu_\mu)}{\Gamma(W^+\rightarrow e^++\nu_e)} \sim \frac{|\sum_i
{-i\rho^{\bf k}_{2,i}\lambda^{\bf
k}_{1,i}}U_{\mu,i}U^\ast_{e,i}|^{2}}{|\sum_i \rho^{\bf
k}_{2,i}\rho^{\bf k}_{1,i}|U_{e,i}|^2|^{2}}\,.
\end{equation}
One can see  $R_{\nu_\mu/\nu_e}\sim O(\frac{m_i^2}{k^2})$\,(the
first term in (\ref{odamplitude}) gives $O(\frac{m_i^4}{k^4})$
contribution; for terms with $\delta(\omega_W - \omega_e +
\omega_i)$\,, we can't find a proper momentum satisfying the
equation of $\omega_W - \omega_e + \omega_i=0$ for on-shell
particles, so we omit their contributions). This is a pure flavor
changing current effect, which contradicts to our starting
Hamiltonian (\ref{cclagran}). It is small for relativistic
neutrinos and vanishes when neutrino is massless/degenerated. And
it is the same order of magnitude for corrections in
\emph{inequivalent vacua model} to the usual Pontecovo's formulas
in (\ref{UinP}). When we go beyond the relativistic limit, the
corrections will be
large, and the flavor changing current effect is also considerable. \\
It happens also in the $\nu_\tau$ family. These off-diagonal decay
modes mean that the definition of weak neutrino states from mixing
fields quantization in the \emph{inequivalent vacua model} cannot
properly describe neutrino interactions. In fact, another
definition of neutrino weak states is on the basis of neutrino
interactions. In our usual knowledge, neutrino weak states are
defined to interact with corresponding charge leptons diagonally
at tree level, just as the Hamiltonian in (\ref{cclagran}). And so
far, the flavors of neutrinos in experiments are also identified
with the signals of corresponding charge leptons. So the emergence
of off-diagonal $CC$ interactions
will spoil the basis of flavor neutrino identification.\\
The problems discussed above also emerge in $NC$ interactions. Let
us discuss the decay of $Z^0$ boson at tree level $Z^0\rightarrow
{\bar \nu_\sigma}+\nu_\rho\,$. Modes with $\sigma=\rho$ indicate
the usual interactions in $SM$. But similar to $CC$ interactions,
modes for different flavors are also nontrivial due to the usage
of the Fock space in the \emph{inequivalent vacua model}\,. But
these flavor changing neutral currents are also forbidden in $SM$
and by experiments.\\
$\bullet$ \emph{Discussions}: In QFT, particles are excitations of
the corresponding fields, but for weak eigenfields, which is the
mixing of mass eigenfields, it is difficult to define the
corresponding quanta. At a glance it looks like that the
\emph{inequivalent vacua model} has overcome this difficulty.
However, the artificial expansions of the weak eigenfields make it
difficult to define an unique Fock space. it is improper to
describe the weak interactions, and inconsistent with the flavor
neutrino definition in weak interactions. The appearance of flavor
changing currents is essential in this model. Its origin is the
anticommutations in  (\ref{commu1}-\ref{commu2t}).

\section{Conclusions}
Physicists want to give an unified description of neutrino
oscillation and neutrino interaction in the framework of QFT. In
the \emph{inequivalent vacua
model}\cite{BV95,BV98,BV02,remarks,space,BV05}\,, they think the
importance of this topic is the \emph{bogliubov} transformation
between the two vacua. In this paper, we compute weak interaction
vertices using the Fock space proposed in their model. From a $CC$
process $W^+\rightarrow e^++\nu_e$\,, we learn that in the
complicated expression of (\ref{ccamplitude}), if $\delta$
functions about energy is explained as energy conservation,
negative energy neutrinos emerge in the process, otherwise this
process violates the principle of energy conservation. We also
compute a flavor changing process $W^+\rightarrow e^++\nu_\mu$ at
tree level and find there is flavor changing current. Estimated
branching ratio of this mode to the standard $W^+\rightarrow
e^++\nu_e$ channel is at the order of $O(\frac{m_i^2}{k^2})$,
which is the same order of the correction to standard Pontecovo's
theory from the \emph{inequivalent vacua model}. Existence of
flavor changing currents will spoil our usual concepts on the
definition of neutrino weak states in neutrino interactions. Only
in the special case of neutrino mass degeneracy (massless limit is
a particular situation of this case), these problems can be
resolved. But the fact of neutrino oscillations has excluded this
case.

\section{Acknowledgments}

Q.Y.L. is grateful to Carlo Giunti for pointing out that the
different $\delta$ functions with respect to $\omega_i$ in
(\ref{odamplitude}) can be considered to be the same and taken out
of the sum. We agree his remark on $\emph{inequivalent vacua
model}$ in ref.\cite{Giunti03}.
\\
The authors would like to thank M.J. Luo and B.L. Chen for useful
discussions. This work is supported in part by the National
Natural Science Foundation of China under grant number 90203002.

\section{Appendix: oscillation effect in weak decay}

If we use real plane waves for  particles. It means the space-time
for the process is infinity, thus one expects that neutrino
oscillation effect will appear in the result. In this case the
processes for (\ref{odamplitude}) and (\ref{ccamplitude})
 are both incoherent superpositions of neutrino mass eigenstate processes with
different energy $\delta$ functions. Under this situation, the
oscillation effect is bigger enough to neglect the
\emph{inequivalent vacua model} effect. We  omit terms with
$\lambda^{\bf k}$, and take $\rho^{\bf k}\simeq1$. We can also
omit dependence of the spinor calculations on neutrino mass for
relativistic case. But dependence of the $\delta$ functions on
neutrino mass $m_i$ can not be neglected in any case. After above
simplification, we can immediately estimate the ratio of the two
processes
\begin{equation}
R_{\nu_\mu/\nu_e}\equiv\frac{\Gamma(W^+\rightarrow
e^++\nu_\mu)}{\Gamma(W^+\rightarrow e^++\nu_e)} \simeq
\frac{\sum_i |U_{\mu,i}U^\ast_{e,i}|^{2}}{\sum_i |U_{e,i}|^{4}}\,.
\end{equation}
By using the approximative tri-bimaximal mixing matrix\cite{HPS},
we obtain an estimated value of the branching ratio which is
$R_{\nu_\mu/\nu_e}\simeq 2/5$\,.
\\
This is exact the averaged (over time) oscillation ratio of
$P(\nu_e \rightarrow \nu_\mu)$ to $P(\nu_e \rightarrow \nu_e)$:
\begin{eqnarray}
\overline{P}(\nu_e \rightarrow \nu_\mu)=\sum_i
|U_{\mu,i}U^\ast_{e,i}|^{2}
\\
\overline{P}(\nu_e \rightarrow \nu_e)=\sum_i |U_{e,i}|^{4}
\end{eqnarray}
\\
The sum of three decay width $W^+\rightarrow e^++\nu_e\,,
W^+\rightarrow e^++\nu_\mu\,$ and $W^+\rightarrow e^++\nu_\tau$
equals the width of $W^+\rightarrow e^++\nu_e\,$ in $SM$. That is
because of the relation of
\begin{equation}
\sum_i
\{|U_{e,i}|^{4}+|U_{\mu,i}U^\ast_{e,i}|^{2}+|U_{\tau,i}U^\ast_{e,i}|^{2}\}=1\,.
\end{equation}
So it doesn't add extra width to the total width of $W^+$ decay.

\end{document}